\documentclass[doublecol]{epl2} %for 2 columns style without line numbers
\usepackage{psfrag,ulem}
\usepackage{amssymb}
\usepackage{amsmath}
\usepackage{color}
\usepackage{hyperref}
\usepackage{mathtools}
\usepackage{color}   %for coloring text
\usepackage{soul}    %for crossing out a line

\newcommand\eq[2]{\begin{equation}\begin{split}#1\end{split}\label{#2}\end{equation}}
\newcommand\fg[3]{\begin{figure}\begin{center}#1\end{center}\caption{#2}\label{#3}\end{figure}}

\title{Critical dynamics of classical systems under slow quench}

\shorttitle{Slow quench dynamics} %Insert here a short version of the title if it exceeds 70 characters

\author{Priyanka \and Kavita Jain}

\institute{Theoretical Sciences Unit, Jawaharlal Nehru Centre for Advanced Scientific Research, Jakkur P.O., Bangalore 560064, India}


\abstract{We study the slow quench dynamics of a one-dimensional nonequilibrium lattice gas model which exhibits a phase transition in the stationary state between a fluid phase with homogeneously distributed particles and a jammed phase with a macroscopic hole cluster. {Our main result is that in the critical region ({\it i.e.}, at the critical point and in its vicinity) where the dynamics are assumed to be frozen in the standard Kibble-Zurek argument, the defect density exhibits an algebraic decay in the inverse annealing rate with an exponent that can be understood using critical coarsening dynamics. However, in a part of the critical region in the fluid phase, the standard Kibble-Zurek scaling 
holds.} 
We also find that when the slow quench occurs deep into the jammed phase, the defect density behavior is explained by the rapid quench dynamics in this phase.}

\begin{document}

\maketitle

%====================================================================================
%Intro
%====================================================================================

\section{Introduction}
\label{Intro}

In the past few decades, extensive studies have been carried out to understand the phase ordering dynamics of classical systems with equilibrium \cite{Bray:1994} and nonequilibrium steady states \cite{Godreche:2001,Godreche:2005,Grosskinsky:2003} when the system is quenched infinitely fast from a disordered state to an ordered one. Slow quench (or annealing) dynamics have also attracted some attention in the recent years, and have been invoked to understand the defect structures in the early universe \cite{Kibble:1976,Kibble:2007} and more generally, to systems exhibiting second order phase transitions \cite{Zurek:1985,Zurek:1996}  in both classical \cite{Huse:1986,Cornell:1992,Vollmayr:1996,Lipowski:2000,Biroli:2010,Suzuki:2009,Krapivsky:2010} and quantum \cite{Laguna:1997,Zurek:2005,Mukherjee:2007,Dziarmaga:2010,Chandran:2012} settings.
Moreover, a number of experiments investigating the relationship between defect density and quench rates have also been performed in a variety of systems such as liquid crystals, superfluid ${^3}$He, superconductors, Bose-Einstein condensates and colloidal systems, see \cite{Campo:2014} for a recent review. 
The defect density at the end of the quench is generally found to decay as a power law in the inverse quench rate (although some systems such as 2D XY model exhibit non-algebraic decay \cite{Jelic:2011,Deutschlander:2015}). 
 
 Much of the  body of work on slow quench dynamics appeals to the Kibble-Zurek argument \cite{Kibble:1976,Zurek:1985} which states that if the control parameter is varied slowly across the critical point, the system stays close to the steady state ({\it adiabatic regime}) until its relaxation time becomes longer than the quench time after which the dynamics  are hypothesised to remain ``frozen" ({\it impulse regime}) until the critical region is crossed.  {Thus the Kibble-Zurek argument describes the quench dynamics before the critical point is crossed. Recent works have elucidated the slow annealing dynamics when the system is quenched far from the critical point to an ordered phase and argued that the defect density is determined by coarsening dynamics \cite{Biroli:2010}.
 %However the dynamical behavior  when the system is quenched slowly {\it in the critical region} has not been investigated much. 
The dynamics in the 2D XY model in which the system is quenched slowly from the high-temperature disordered phase to the low-temperature critical phase have also been studied \cite{Jelic:2011}. However, a complete study of a system with isolated critical point in which one can examine the change in the dynamical behavior when a slow quench occurs below, at and above the critical point has not been carried out.} Moreover, all the studies mentioned above deal with systems with equilibrium steady state (but, see \cite{Karevski:2016} that considers a nonequilibrium quantum system).

Here we consider the slow quench dynamics of the jamming transition that occurs in diverse settings such as vehicular traffic  \cite{Chowdhury:2000}, cellular traffic \cite{Leduc:2012} and granular media \cite{Liu:1998}. We study a 
classical nonequilibrium system in one dimension which shows a jamming transition in the stationary state \cite{Evans:2006}. The steady state of this model is known exactly \cite{Evans:2005}, and some results for the coarsening dynamics \cite{Godreche:2003,Grosskinsky:2003} and stationary state dynamics \cite{Priyanka:2016} have also been obtained. However the dynamics of this model under slow annealing have not been studied and here we address this problem using numerical simulations. We find that  the standard Kibble-Zurek scaling explains our results {in a part of the critical region} but close to the critical point and for quenches deep in the jammed phase, the defect density decay can be understood using the corresponding results for rapid quench dynamics \cite{Biroli:2010}.

%====================================================================================
%
%====================================================================================

\section{Models}

We consider a one-dimensional lattice with periodic boundary having $L$ sites and $N$ particles. 
The total number of particles in the system is conserved. Due to hard-core interactions between the particles, each site can have at most one particle. As shown in Fig.~\ref{domain}, a particle hops to  its right 
empty neighbor with a rate $u(n)$ where $n$ is the number of holes in front of it ({\it unidirectional model}).

It is useful to map this model to a  zero range process (ZRP) with ${\cal{L}}=N$ sites and ${\cal{N}}=L-N$ mass units by considering particles in the lattice gas model as sites and holes as mass clusters \cite{Evans:2005}. Thus in ZRP, a site can
contain any number of particles and a particle hops out of a site containing $n$ particles to its left neighbor  with a rate $u(n)$, see Fig.~\ref{domain} for an illustration. For an infinitely large system in the stationary state, the probability that a site contains $n$ particles is given by \cite{Evans:2005}
\begin{equation}
  p(n)=\omega^{n}\frac{f(n)}{g(\omega)}~,
  \label{mass-dis}
 \end{equation}
 where 
 \begin{equation}
  f(n)=\prod_{k=1}^{n}\frac{1}{u(k)}  ~,~n \geq 1.
 \end{equation}
 In the above equation, $g(\omega)=\sum_{n=0}^{\infty}\omega^{n} f(n)$ is the generating function of $f(n)$ 
 and the fugacity $\omega$  is calculated from the fugacity-density relation, 
 \begin{equation}
  \varrho= \frac{\cal N}{\cal L}=
    \omega \frac{\partial \ln g(\omega)}{\partial \omega} ~.
\label{fug_density}
 \end{equation}
From the above equation, it can be seen that the fugacity is an increasing function of the density $\varrho$. For certain choices of $u(n)$, the fugacity reaches its maximum value (given by the radius of convergence of the generating function $g(\omega)$) at a finite density $\varrho_c$.  Then for  $\varrho > \varrho_c$, the excess density $\varrho-\varrho_c$ resides at a single site while for $\varrho < \varrho_c$, each site supports a density $\varrho$ \cite{Evans:2005}. 
 
 For the hop rate
 \begin{equation}
  u(n)=1+\frac{b}{n}~,~n > 0 ~,~
  \label{hoprate}
 \end{equation}
 the ZRP shows the phase transition described above in the $\varrho-b$ plane for $b > 2$. Correspondingly, 
the lattice gas model exhibits a phase transition at the critical point given by
  \begin{equation}
   b_c=\frac{2-\rho}{1-\rho} ~,
   \label{bceqn}
  \end{equation}
  where $\rho=N/L$ is the total particle density. For  $b < b_c$, the typical hole cluster length is  of order unity (fluid phase) while for $b > b_c$, a macroscopically long hole cluster coexists with gaps that are power law distributed as $n^{-b}$ (jammed phase) \cite{Evans:2005}. It has been shown that close to the critical point, the static correlation length $\xi \sim(b_c-b)^{-\nu}$ where the exponent $\nu$ varies continuously with $b_c$ when  $2 < b_c < 3$ but it is a constant otherwise \cite{Priyanka:2014}: 
  \begin{equation}
  \label{nures}
  \nu=
  \begin{cases}
  (b_c-2)^{-1} &,~~ 2 < b_c < 3 \\
  1 &,~~ b_c \geq 3~.
  \end{cases}
  \end{equation}
The stationary state dynamics have also been studied and it has been found that at the critical point, the steady state density fluctuations decay on a time scale that grows as $L^z$ where the dynamic exponent $z=3/2$ \cite{Priyanka:2016}. 

In the following, we  also consider a {\it bidirectional model} in which the particle first chooses either the left or the right neighbor  with equal probability and then hops with a rate that depends on the vacancies in the chosen direction provided the target site is empty.  The steady state obeys detailed balance and is the same as that in the unidirectional model  \cite{Evans:2005}. As a result, the correlation length exponent $\nu$ is given by (\ref{nures}); however, the dynamic exponent $z=2$ in this case \cite{Priyanka:2016}.

%###################### Fig 1 ###################################################
\fg{
\includegraphics[width=0.48\textwidth]{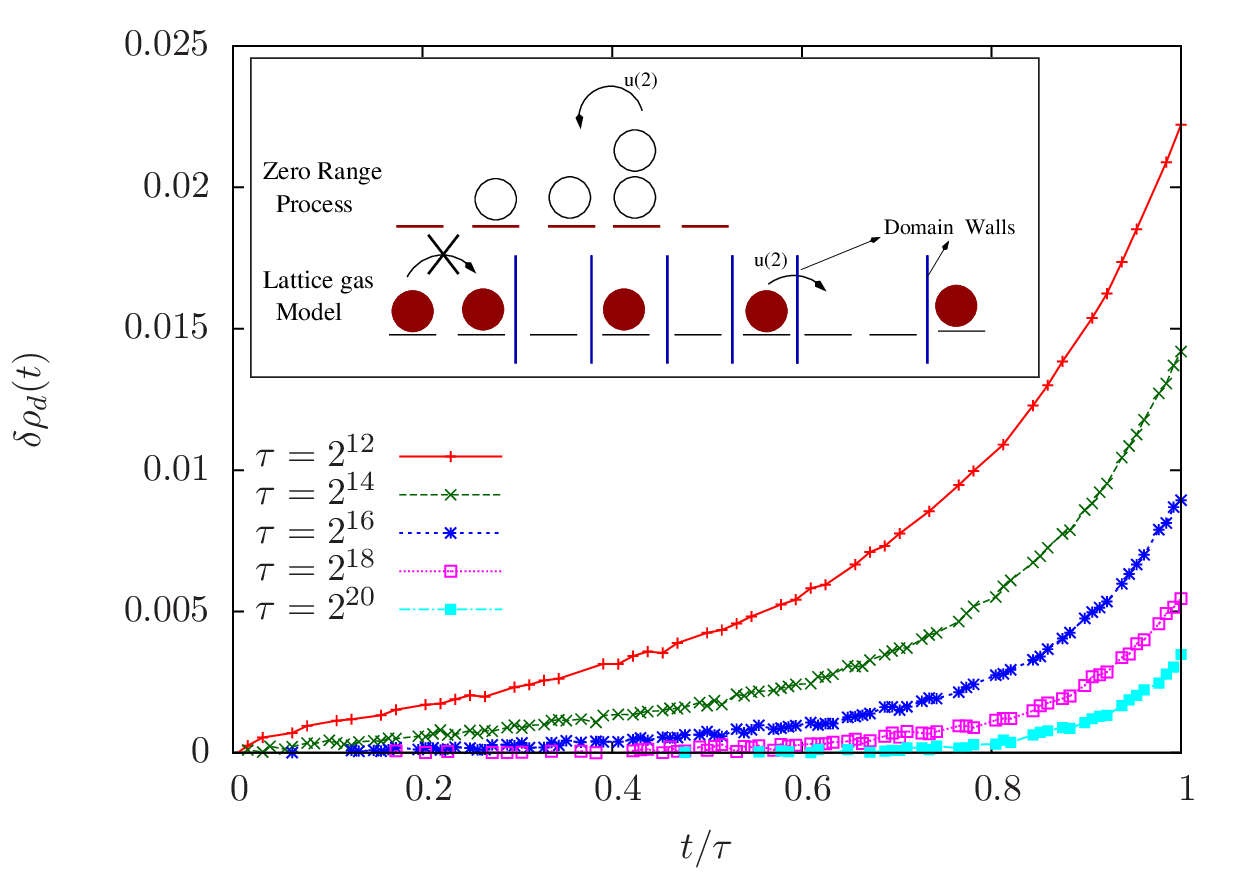}}
{Inset shows the unidirectional lattice gas model studied in this article and the related zero range process. Main figure shows the dynamics of the excess defect density defined in (\ref{dw}) for $b_\tau=b_c=2.3$ in the unidirectional model.}{domain}
%#########################################################################
 
To study the slow quench dynamics, we introduce time-dependence in the hop rates and write 
  \begin{equation}
   u(n,t)=1+\frac{b(t)}{n}~,~n > 0 ~,~
  \end{equation}
 where, for simplicity, we work with linearly varying $b$ given by 
\begin{equation}
 b(t)= \frac{b_\tau t}{\tau}~,~ 0 \leq t \leq \tau ~.
 \label{quenchpro}
\end{equation}
The quench protocol was carried out by changing the parameter $b$ from zero (in the fluid phase) to a final value $b_\tau=b_c$ (critical point) and $2 b_c$ (jammed phase) keeping the density fixed at $\rho_c$ given by (\ref{bceqn}). Our quantity of interest is the domain wall density (which is the interface between the particle and hole) in the lattice gas model. From the inset of Fig.~\ref{domain}, we see that the number of domain walls is equal to two times the number of occupied sites in the ZRP. For finite inverse quench rate $\tau$, as the system is far from its steady state and has more domain walls than in the stationary state, we consider the excess defect density given by 
\eq{
\delta \rho_{d}(t)=2\rho_c [p(0, t)-p(0)] ~,}{dw}
where $p(0,t)$ is the probability that a site is empty at time $t$ in ZRP when the parameter $b$ is time-dependent and $p(0)$ is given by (\ref{mass-dis}). In Monte-Carlo simulations of the models described above, we measured $\delta \rho_{d}(t)$ for system sizes in the range $15000- 20000$ and averaged the data over $2000-4000$ independent initial conditions.

%====================================================================================
%Results
%====================================================================================
\section{Results}  When the quench rate  $\tau^{-1}$ is very small, the parameter $b$ changes very slowly  allowing the system to relax to the stationary state. But for faster quench, the system is farther from the steady state. Indeed, as Fig.~\ref{domain} shows, the excess defect density $\delta \rho_{d}(t)$  decreases with increasing $\tau$. Our objective here is to understand how  
$\delta \rho_{d}(\tau)$ decays with $\tau$ when the system is quenched slowly to $b_\tau$.  

%###################### Fig 2, 3 ###################################################
\fg{
\includegraphics[width=0.48\textwidth]{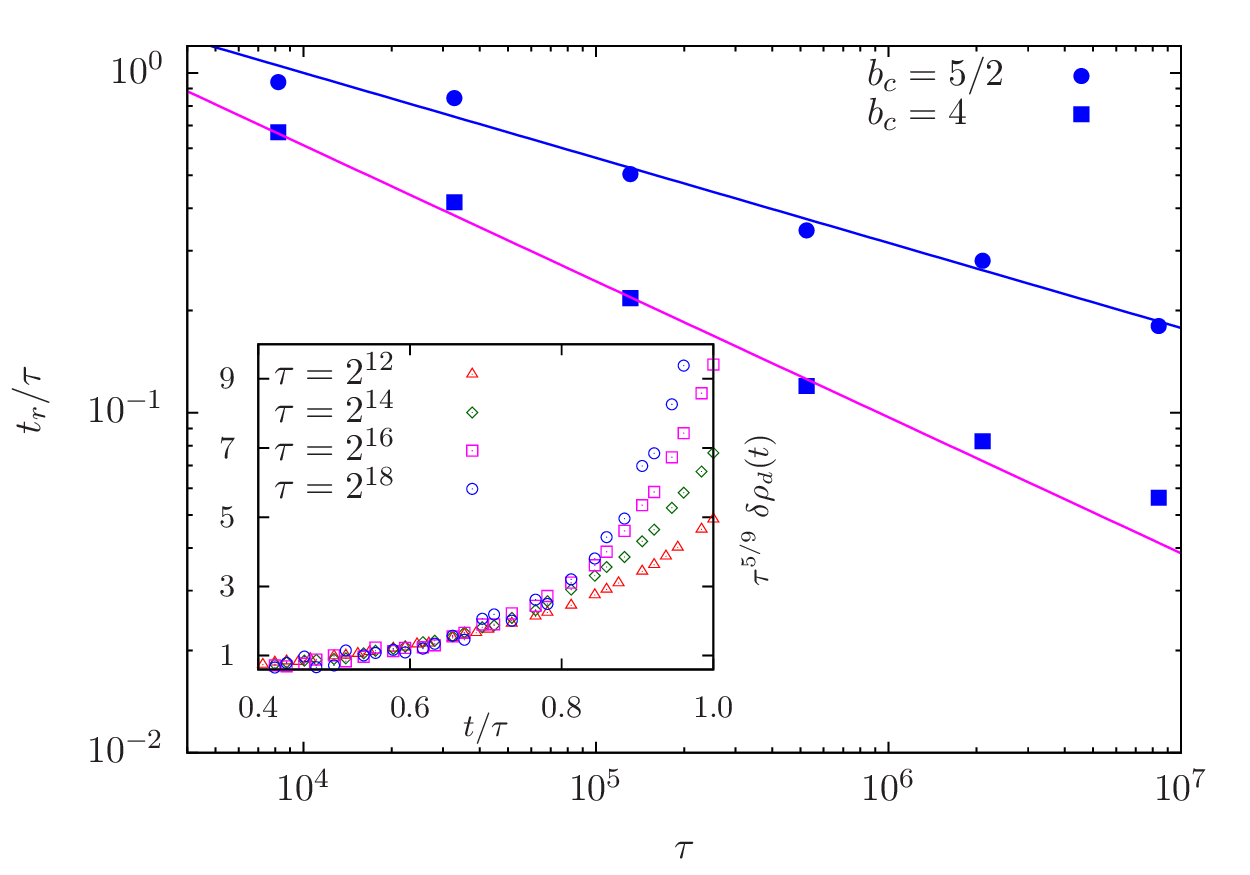}}{Main figure shows that the remaining time $t_r$ to the critical point obeys the Kibble-Zurek prediction  (\ref{th}) for the unidirectional model when $b_\tau=b_c$. Inset shows the collapse of excess defect density  with the Kibble-Zurek scaling (\ref{KZDD}) for time $t_* < t < \tau$ when the system is quenched to $b_\tau=b_c=2.3$.}{fgtcap}

\fg{
 \includegraphics[width=0.47\textwidth]{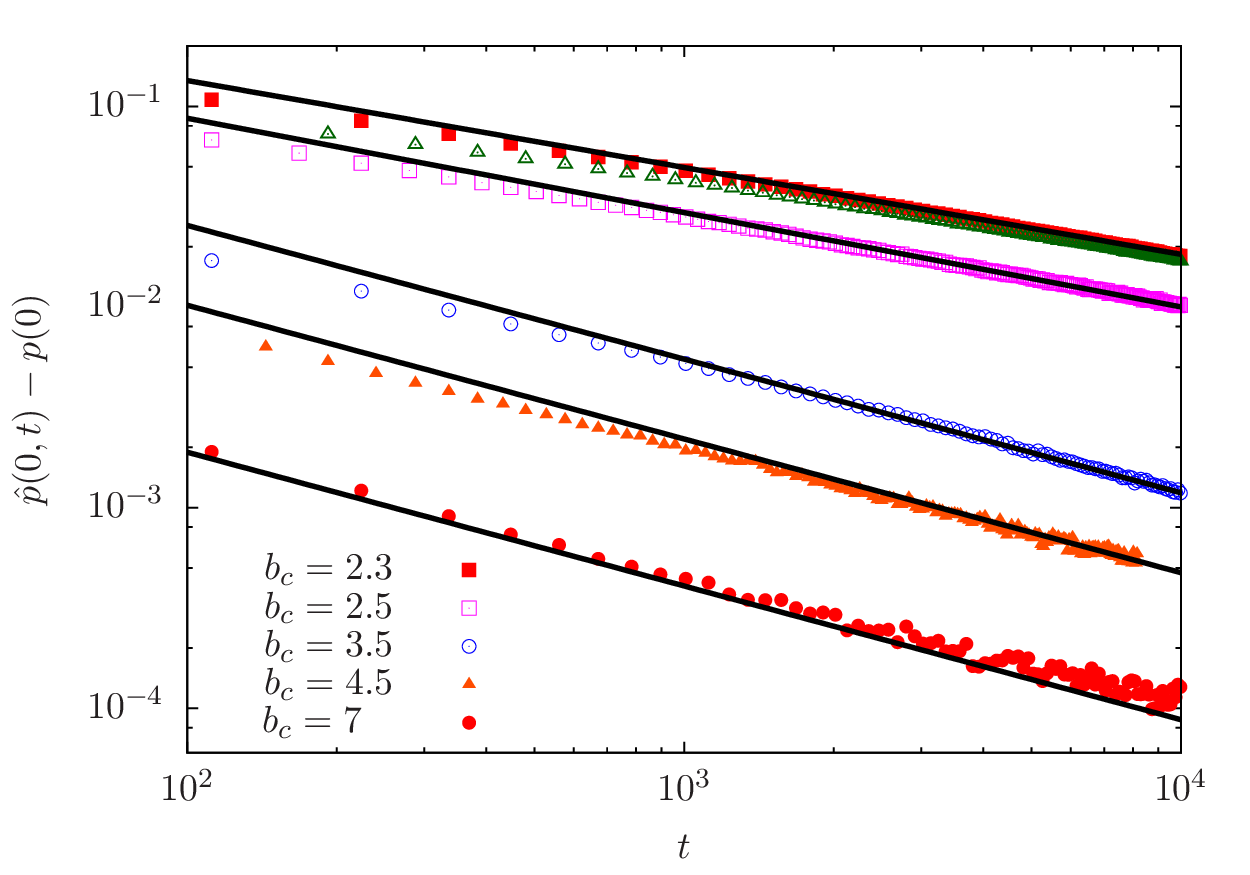}}
 {{Power law decay of the probability ${\hat p}(0,t)-p(0)$ with time after a fast quench to the critical point starting from $b=0$ at density $\rho_c$ in the unidirectional model. The lines show the scaling (\ref{alpha}), and the triangular open  symbols show the numerical data when the system is quenched instantaneously to $b_c=2.3 $ from the initial value one.}}{CoarsenAS}
 
 \fg{
 \includegraphics[width=0.47\textwidth]{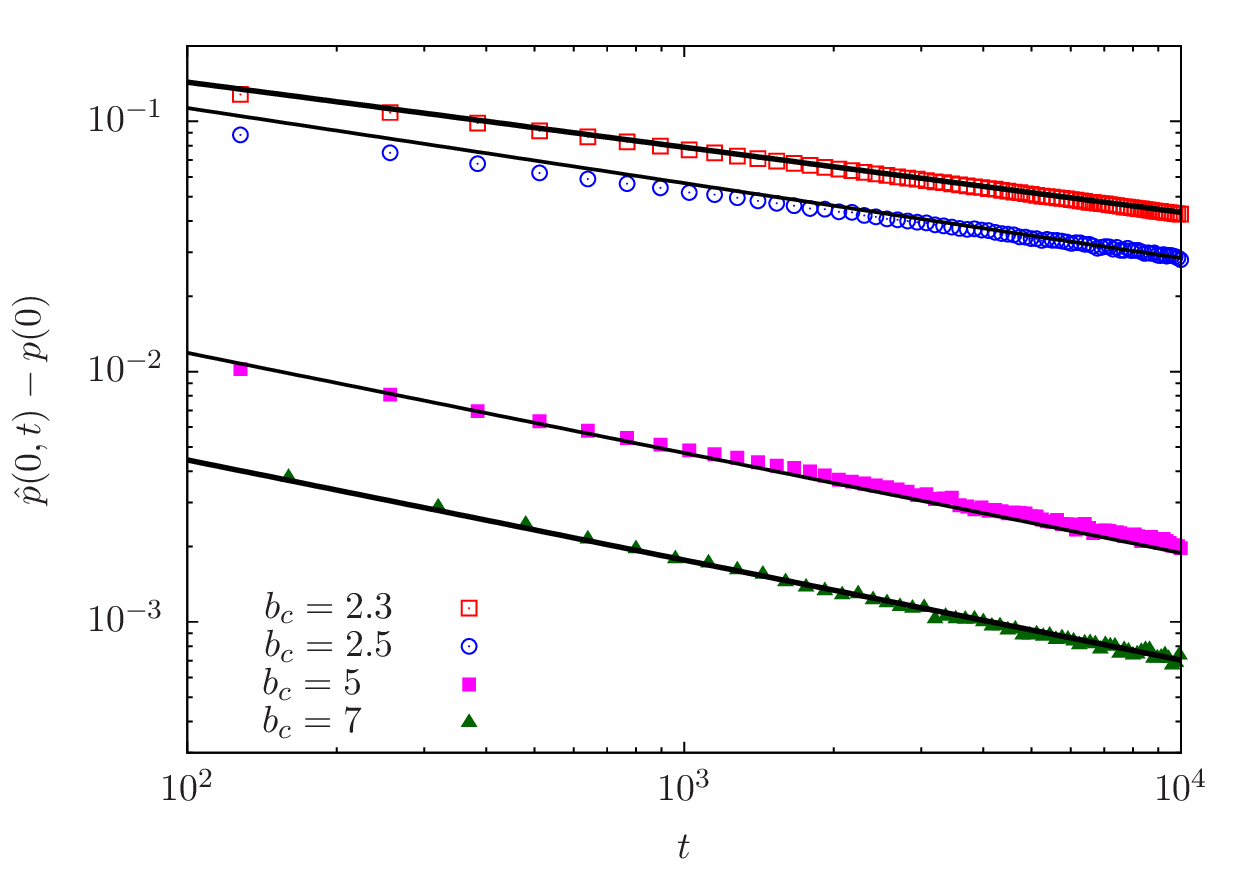}}
 {{Power law decay of the probability ${\hat p}(0,t)-p(0)$ with time after a fast quench to the critical point starting from $b=0$ at density $\rho_c$ in the bidirectional model. The lines show the power law decay with exponent given by (\ref{alpha}).}}{CoarsenS}

%###################### Fig 2, 3,4###################################################

\noindent{\it {Dynamics in the fluid phase close to the critical point:}}
When the system is far from the critical point, it relaxes quickly. 
But as the critical point is approached, the relaxation time increases and at time $t_*$,  the remaining time $t_r=\tau-t_*$ to reach the critical point becomes comparable to the relaxation time in the stationary state which can be expressed as \cite{Kibble:1976,Zurek:1985} 
\eq{
\tau-t_{\ast} \sim \xi_*^{z} \sim (b_c-b(t_*))^{-z \nu} \sim \left(1- \frac{t_*}{\tau} \right)^{-z \nu} ~,}{th}
so that $t_r \sim \tau^{\frac{z \nu}{1+ z \nu}}$. As the system falls out of equilibrium at $t_*$,  in numerical simulations, we picked the time $t_{\ast}$ to be the one where the excess domain wall density is $10^{-3}$ and found the time $t_r$. Using the exponents $\nu$ and $z$ quoted in the last section, we find that for the unidirectional model, the time $t_r \sim \tau^{3/(2 b_c-1)}$ for $2< b_c <3$ and as $\tau^{3/5}$ for $b_c \geq 3$ which is in good agreement with the numerical data in the main panel of Fig.~\ref{fgtcap}. (We have checked that our scaling results are not affected if $t_*$ is determined by the criteria that the excess defect density $\lesssim 10^{-3}$).

Assuming that the system does not evolve after time $t_*$, the defect density {\it after} crossing the critical point  is posited to be $\delta \rho_{d}(\tau)\sim \xi_{\ast}^{-1}\sim \tau^{-\nu/(1+\nu z)}$ \cite{Kibble:1976,Zurek:1985}.
However, we find that the Kibble-Zurek scaling works well for times smaller than $\tau$ but not at or after the critical point is crossed (see below). Our simulation results for the unidirectional model shown in the inset of Fig.~\ref{fgtcap} for a quench to the critical point demonstrate that the excess defect density is of the form,
\begin{equation}
  \label{KZDD}
  \delta \rho_d^{(\textrm{uni})} (t) =
  \begin{cases}
\tau^{-\frac{2}{2 b_c-1}} {f_1(t/\tau)} &,~~ 2 < b_c < 3 \\
  \tau^{-2/5} {f_2(t/\tau)} &,~~ b_c \geq 3 ~,
  \end{cases}
  \end{equation}
 where $f_1(x)$ and $f_2(x)$ are the scaling functions, and $t_* < t < \tau$.  We also find that the above scaling form breaks down at  times of order $t_*$. 

%###############################Fig 5,6#########################################################

\fg{
\includegraphics[width=0.48\textwidth]{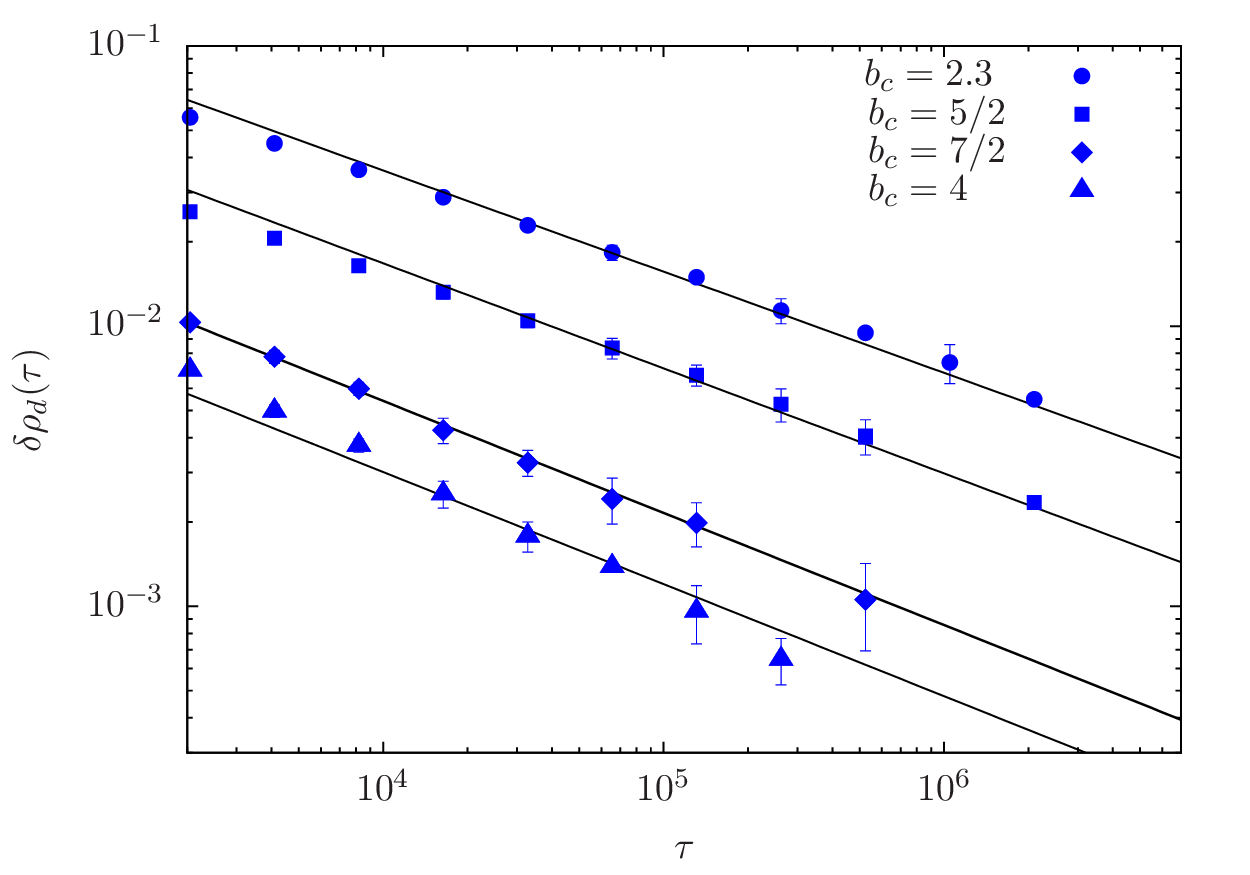}}{Decay of the excess defect density when the system is quenched slowly to the critical point ($b_\tau=b_c$) in the unidirectional model. Our scaling prediction (\ref{PJDD}) is compared with numerical data for several values of $b_c$ and the errorbars for some representative points are also shown. {The data for $b=2.3$ is multiplied by a factor $2$ in order to show all the plots in the same figure.}}{asymm}

\fg{
\includegraphics[width=0.48\textwidth]{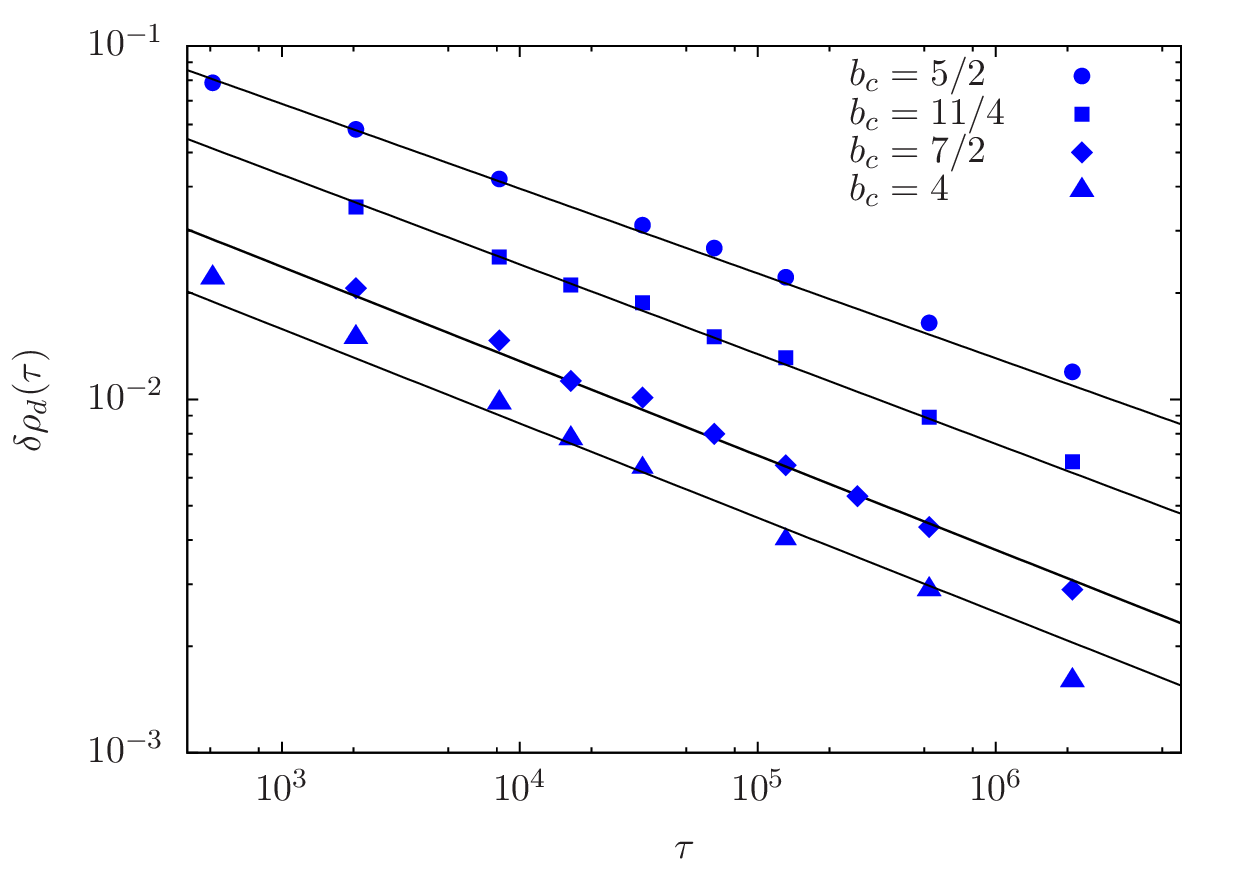}}{Decay of the excess defect density when the system is quenched slowly to the critical point ($b_\tau=b_c$) in the bidirectional model. Our scaling prediction (\ref{PJDDs}) is compared with numerical data for several values of $b_c$; the errorbars in this case are smaller than the point size. {The data for $b=2.5$ is multiplied by a factor $1.5$ in order to show all the plots in the same figure.}}{symm}

%#################################Fig 5, 6#######################################################

\noindent{\it Dynamics at the critical point:} Close to the critical point, the system undergoes critical coarsening during the time interval $t_r$ \cite{Biroli:2010}. We therefore turn to a discussion of the fast quench dynamics of the ZRP in which the system initially in the fluid phase is quenched instantaneously to the critical point. To distinguish between the quantities obtained using slow and fast quench, in the following, we use ${\hat ~}$ to refer to quantities obtained using the fast quench protocol.

In \cite{Godreche:2003}, critical coarsening dynamics of the ZRP have been investigated in mean-field geometry and in one dimension. In the latter case, numerical simulations indicate that a measure of the domain length grows with time as $t^{1/{\hat z}}$ with the coarsening exponent ${\hat z}=3~(5) $ for unidirectional (bidirectional) model and a scaling argument suggests that the probability 
$\delta {\hat p}(0,t) \sim t^{-{\hat \alpha}}$ with the exponent ${\hat \alpha}=(b-2)/{\hat z}$ for $b>3$ (see (35) of \cite{Godreche:2003}). While our numerical results for ${\hat z}$ are in agreement with those of \cite{Godreche:2003}, the exponent ${\hat \alpha}$ for the distribution of empty sites is not consistent with that in \cite{Godreche:2003}. Instead, our numerical results shown in Figs.~\ref{CoarsenAS} and \ref{CoarsenS} suggest that  the exponent $\hat \alpha$ varies with $b_c$ for $2 < b_c < 3$ but it is a constant otherwise, and a best regression fit gives
\begin{equation}
  \label{alpha}
  {\hat \alpha}{\approx}
  \begin{cases}
 (b_c-1)/{\hat z}  &,~~ 2 < b_c < 3 \\
  2/{\hat z} &,~~ b_c \geq 3~.
  \end{cases}
  \end{equation}
%{An understanding of these exponents is currently }
%In the stationary state, using the fact that the mass distribution $p(n) \sim n^{-b}$ at the critical point and extreme value theory \cite{Sornette:1964}, it can be seen that in a system of size $L$, the largest mass is of the order $L^{1/(b-1)}$ for $b > 2$. These considerations also show that 
%\eq{
%{p}(0,L)-{ p}(0) \sim \int_{L^{\frac{1}{b-1}}}^\infty dn~ n^{-b} \sim L^{-1}~,
%}{p0}
%and the mass fluctuation $\ell$ grows as \cite{Evans:2006}
%\begin{equation}
%  \label{xi}
%  \ell \sim
%  \begin{cases}
%   L^{1/(b-1)}&,~~ 2 < b < 3 \\
%  \sqrt{L} &,~~ b \geq 3 ~.
%  \end{cases}
%  \end{equation}
%However, as mentioned above, on rapid quench, the mass fluctuation ${{\hat \ell} (t)} \sim t^{1/{\hat z}}$ \cite{Godreche:2003} on using 
%which in (\ref{xi}) yields a relationship between time and system size thence leading to $\delta {\hat p}(0,t) \sim t^{-{\hat \alpha}}$ with ${\hat \alpha}$ given by (\ref{alpha}).  

Using the above results and recalling that the coarsening process initiates at time of order $t_*$, we obtain $\delta \rho_d (\tau) \sim t_r^{-{\hat \alpha}} \sim \tau^{-z \nu {\hat \alpha}/(1+ z \nu)}$ when the system is quenched to the critical point. Here we have ignored the dependence on the quench depth ({\it i.e.}, $b_c-b(t_*)$) since our numerical results  in Fig.~\ref{CoarsenAS} suggest that the long time dynamics are independent of it. More explicitly, for the unidirectional model, we have 
\begin{equation}
  \label{PJDD}
  \delta \rho_d^{(\textrm{uni})} (\tau) \sim 
  \begin{cases}
\tau^{-\frac{b_c-1}{2 b_c-1}} &,~~ 2 < b_c < 3 \\
  \tau^{-2/5}&,~~ b_c \geq 3~,
  \end{cases}
  \end{equation}
 while, for the bidirectional model, we get
 \begin{equation}
  \label{PJDDs}
  \delta \rho_d^{(\textrm{bi})} (\tau) \sim 
  \begin{cases}
\tau^{-\frac{2 (b_c-1)}{5 b_c}} &,~~ 2 < b_c < 3 \\
  \tau^{-4/15}&,~~ b_c \geq 3~.
  \end{cases}
  \end{equation}
The above predictions for the excess defect density are consistent with the numerical results shown in Figs.~\ref{asymm} and \ref{symm} when $b_\tau=b_c$. We have also checked that the above scaling predictions hold when the system is quenched in the vicinity of $b_c$.

%##################################Fig 6########################################################

\fg{
\includegraphics[width=0.48\textwidth]{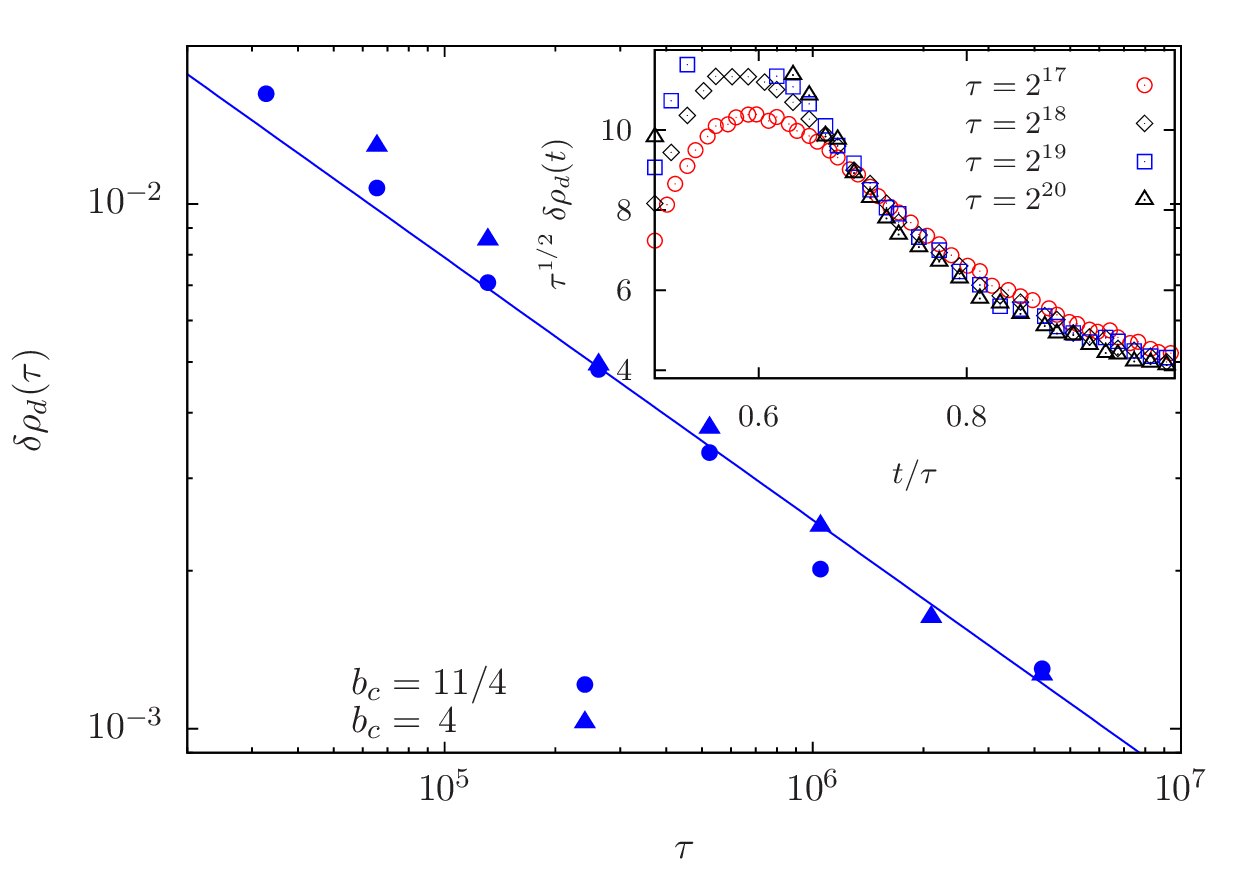}}{Decay of the excess defect density in the unidirectional model when the system is quenched slowly to $b_\tau=2b_c$. The lines in the main figure show (\ref{LCDD}) and the data collapse with the coarsening exponents in the jammed phase is shown in the inset for $b_\tau=2 b_c, b_c=2.3$.}{beyond}

%#################################Fig 6#########################################################

\noindent{\it {Dynamics deep in the jammed phase:}} For $b_\tau \gg b_c$, the system undergoes coarsening in the jammed phase after the time $t_c+t_r$ where $b(t_c)=b_c$ \cite{Biroli:2010}. As the time scale $t_c \sim \tau$ but $t_r$ is sublinear in $\tau$,  the time left until the quench is of order $\tau$.  
{Thus we expect the defect density to simply decay as 
\begin{equation}
\label{LCDD}
\delta \rho_d(\tau) \sim \tau^{-1/{\hat z}_O} ~,~ b_c > 2~,
\end{equation}
where the coarsening exponent in the ordered phase, ${\hat z}_O=2$ (${\hat z}_O=3$) for unidirectional (bidirectional) model \cite{Grosskinsky:2003,Godreche:2003}.} Figure~\ref{beyond} shows that our numerical results for $b_\tau=2 b_c$ are consistent with the above scaling prediction. Moreover, the excess defect density $\delta \rho_d(t)$ is of a scaling form similar to (\ref{KZDD}) as attested by the data collapse shown in the inset of Fig.~\ref{beyond}.

%====================================================================================
%Conclusions%====================================================================================
\section{Conclusions} Slow quench dynamics have been studied extensively when the control parameter is changed across the critical point of a second order phase transition in equilibrium systems \cite{Zurek:1985,Zurek:1996}, and recent works have considered slow quenches {deep into the ordered phase \cite{Biroli:2010} or the critical phase \cite{Jelic:2011}. Here we have performed, to our knowledge, the first quantitative study of the slow quench dynamics when a nonequilibrium system with isolated critical point is quenched in the critical region.}

The Kibble-Zurek argument assumes that in the critical region, the dynamics are frozen since the relaxation times are much longer than the time available. {As shown in the inset of Fig.~\ref{fgtcap}, this argument indeed holds when the system is close to the critical point; more precisely, we find that the Kibble-Zurek scaling is obeyed over a time window 
$\left[t_*, f t_*\right], f < 1$, when the system is quenched to the critical point. However, it does not apply during $f t_* < t < \tau$ (and similarly, in the symmetric time window beyond the critical point in the jammed phase). Instead, we find that the defect density decays as a power law although with an exponent smaller than or equal to the Kibble-Zurek prediction, see Figs.~\ref{asymm} and \ref{symm}.} While the dynamics outside the critical region involve only the stationary state dynamic exponent ($z$) and the deep quench in the ordered phase is determined by the coarsening exponent (${\hat z}_O$), the critical point quench dynamics are more complex involving both static fluctuations and critical coarsening. 

For the class of models considered here, a comparison of the decay exponents shows that the excess defect density decays faster in the unidirectional model which has a nonequilibrium steady state than in the bidirectional model with equilibrium steady state.  We also obtain continuously varying exponents for $2< b_c < 3$ where the mass fluctuations are anomalous but constant otherwise \cite{Evans:2005}. 

A detailed exploration of other nonequilibrium and equilibrium systems with critical point annealing should inform us better about the dynamics in the impulse regime and is a goal for the future.

 \section{Acknowledgment}
 The authors thank Diptiman Sen, Amit Dutta and Tanmoy Pal for helpful discussions. Priyanka acknowledges the University Grants Commission for financial support through a research fellowship. We also thank two anonymous reviewers for useful comments. 

%\input{Draft3.bbl}

%--------------------------------------------------------------

\end{document}